\documentclass[draft]{iopart}
\usepackage{iopams}  

\usepackage[final]{graphicx}
\usepackage{graphicx}
\usepackage{epsfig}

\usepackage[normalem]{ulem}
\usepackage[usenames]{color}

\begin{document}

\title[]{ The exchange bias phenomenon in uncompensated interfaces: Theory and Monte Carlo simulations}

\author{O V Billoni$^1$, S A Cannas$^1$ and F A Tamarit$^1$}

\address{$^1$Facultad de Matem\'atica, Astronom\'{\i}a y
F\'{\i}sica, Universidad Nacional de C\'ordoba and Instituto de
F\'{\i}sica Enrique Gaviola  (IFEG-CONICET), Ciudad Universitaria,
5000 C\'ordoba, Argentina.}
\ead{billoni@famaf.unc.edu.ar}
\ead{cannas@famaf.unc.edu.ar}
\ead{tamarit@famaf.unc.edu.ar}

\begin{abstract}
We performed Monte Carlo simulations in a bilayer system composed of 
two thin films, one ferromagnetic (FM) and the other antiferromagnetic (AFM). Two lattice structures for the films were considered: 
simple cubic $sc$ and a body center cubic $bcc$. We imposed an uncompensated interfacial
spin structure in both lattice structure; in particular we emulated a FeF$_2$-FM system in the case of the $bcc$ lattice.
Our analysis focused  on the incidence of the interfacial strength interactions between the films, $J_{eb}$, and the effect of thermal 
fluctuations on the  bias field, $H_{EB}$. We first performed Monte Carlo simulations on a microscopic model based on classical 
Heisenberg  spin  variables. To analyze the simulation results we also introduced a simplified model that assumes coherent rotation 
of spins located on the same layer parallel to the interface. We found that, depending on the AFM film anisotropy to exchange ratio, 
the bias field is either controlled by the intrinsic pinning of a domain wall parallel to the interface or by the stability of the 
first AFM layer (quasi domain wall) near the interface.
\end{abstract}

\pacs{75.70.-i, 75.60.Jk, 75.70.Cn}

\noindent{\it Keywords}: Thin films, magnetic properties, magnetization reversal mechanisms.  
\submitto{\JPCM}
\maketitle

\section{Introduction}

Exchange bias (EB) is an ubiquitous magnetic phenomenon that usually appears
when two different magnetic media are in contact. Although EB can be observed
in a large variety of non-homogeneous magnetic materials \cite{Nogues05PR, Nogues99JMMM},
in this work we will focus on the case of a bilayer system composed of two films, one ferromagnetic 
(FM) and the other antiferromagnetic (AFM).

Assuming that the Curie Temperature $T_C$ of the ferromagnetic material is larger than
the N\`eel Temperature $T_N$ of the antiferromagnetic one, and that the two films are magnetically
coupled by exchange interactions, an unusual hysteresis phenomenon can be observed.
If such a system is cooled down below $T_N$ in the presence of an external applied magnetic
field $H_{CF}$ the hysteresis loops of the FM material evidences three important anomalies
when compared with the loop of the single ferromagnetic film. First,
a shift in the loop appears, characterized by a new center called the bias field $H_{EB}$.
This shift is due to the unidirectional anisotropy induced at the interface.
Second,  the width of the loop usually increases. Finally, the loop also loses its symmetry.
As temperature increases, the bias field $H_{EB}$ goes to zero at certain blocking
temperature $T_B$, with $T_B < T_N$, restoring the normal hysteresis loop of the
isolated ferromagnet.

Although this phenomenon was reported for the first time in 1956 \cite{Meiklejohn56PR} and
despite the huge theoretical and experimental effort devoted to understanding its origins,
there are still many controversial points concerning the underlying mechanisms responsible
for such unusual hysteresis anomalies
\cite{Nogues05PR,Nogues99JMMM, Kiwi01JMMM,Berkowitz99JMMM, Radu08}.
In particular, these controversies are in part related to the fact that EB has been observed in
a great diversity of magnetic system, including for instance spin glasses, intrinsic
inhomogeneous and nanoparticle systems, as well as the bilayered system  analyzed
in this paper.  Beyond the theoretical interest, this phenomenon is also relevant because of its
technological applications--for instance, in the design of
 magnetic sensor and magnetic recording media devices \cite{Nogues99JMMM}, among many others.

As regards the case of a bilayered FM/AFM system, the spin structure at the interfacial planes
is a main issue in developing the understanding of the EB phenomenon. In particular, AFM interfaces
can be roughly classified as {\em compensated} or {\em uncompensated}, depending on whether the
nearest AFM plane parallel to the interface have zero net magnetization or not, respectively. 
Most of the earlier models that explain EB assume an uncompensated interfacial spin
structure \cite{Kiwi01JMMM}, even when this requirement is not always fulfilled in experiments.
Actually, EB can be observed also in compensated interfaces, but in this case the
existence of uncompensated domains has shown to be fundamental for the appearance of
the hysteresis shift \cite{Takano97PRL}. Furthermore, fully uncompensated
interfaces can manifest a weaker EB field when compared with partially uncompensated or even
compensated interfaces.
In fact, experiments carried out by Moran {\it et al.} \cite{Moran98APL} and Nogu\'es
{\it et al.} \cite{Nogues99PRB} on Fe films grown over FeF$_2$ single crystals  cut along
different orientations, showed that $H_{EB}$ is larger when the interface is compensated
((110) plane) in comparison with the uncompensated case ((010) plane).
This effect is supposed to be associated with spin re-arrangement at the interface
\cite{Nogues99JMMM,Nogues99PRB} since a similar behavior was found when the AFM is a single
crystals or a thin film.

A key point for the understanding of the EB phenomenon on uncompensated interfaces is the effect of the  
variation of the exchange coupling between interface layers  on the EB field. While it is difficult 
to control this quantity at the experimental level, this problem can be handled easily using Monte Carlo 
simulations based on microscopic models. In addition, this methodology allows a detailed description of 
the interfacial spin structure together with the incorporation of thermal  fluctuations, which are relevant for
the stability and therefore the appearance of the EB phenomenon. For instance, thermal effects are necessary 
to explain the widening of the hysteresis loop close to the blocking 
temperature \cite{Scholten05PRB,Leighton02JAP, Ohldag06PRL}.
In this sense, numerical studies at the  micromagnetic \cite{Suess02IEEE,Suess03PRB,Dorfbauer05JMMM}
and Monte Carlo simulations levels \cite{Billoni06PB,Spray06JPDAP, Lederman04PRB,Misra04JAP,Mitsumata03PRB,Sakurari03JAP,Nowak02PRB,Nowak02JMMM} have proved to be very useful tools for modeling realistic systems.
On the other hand, the continuous approximation assumed in micromagnetic based model breaks 
down in highly anisotropic materials like FeF$_2$ antiferromagnetic compounds. Discretization could give rise 
to different energy barriers with the consequent thermal activated effects \cite{Kim05PRB}. Hence, atomic scale based  
models could be crucial for getting an appropriated description of the magnetic properties.

In this paper we analyzed the EB phenomenon in a FM--AFM bilayer system with an uncompensated interface. 
In section \ref{background} we  summarize the existing theoretical background, discussing
the phenomenology of EB system in the frame of two of the most relevant models. In section \ref{simulations} 
we introduce a microscopic model for the bilayered system, describe the simulation protocol and show
our numerical results. 
In order to interpret  the results of the previous section
we introduce in section \ref{model} a generalization  of Meiklejohn-Bean model, which allowed us to analyze the 
role of the AFM  layers in the EB phenomenon. In section \ref{conclu} we summarize and discuss the results.

\section{Theoretical Background}

\label{background}
In order to analyze the role of the strength of the interface exchange interaction $J_{eb}$ in the behavior
of bias field $H_{eb}$, let us discuss first the following question: what happens with the order of the AFM as we
invert the orientation of the FM magnetization by applying an opposite magnetic field $h$? We assume that
the system has already reached thermal equilibrium at certain temperature $T$ below $T_B$, in such a way that,
if $J_{eb}$ were zero, both films would have achieved an ordered state. Since $T_C > T_N$, we assume $|J_F| > |J_A|$
where $J_F$ and $J_A$ are the exchange interactions of the FM and AFM, respectively. If $J_{eb}$ is small
enough ($J_{eb} << J_A$) the spins in the AFM will remain almost insensitive to the rotation of the global
magnetization of the FM film. In this case the Meiklejohn-Bean model \cite{Meiklejohn62JAP} predicts a linear
dependency of the bias field $H_{eb}$ on the value of $J_{eb}$:
\begin{equation}
\label{eq2}
H_{EB}  \propto \frac{J_{eb}}{L_{FM}},
\end{equation}
where $L_{FM}$ is the thickness of the FM film.

At the other extreme, when $J_{eb}$ is large enough, the rotation of the magnetization would induce the creation of
a domain wall (parallel to the interface) in the AFM films, at least for small enough values of $K_A$. Once a perfect
domain wall has been formed, any increase of $J_{eb}$ will not alter the value $H_{eb}$.
This phenomenology is captured by  the model of Mauri {\it et al.}  \cite{Mauri87JAP} (from now on the MSBK-model) 
when the anisotropy of the FM film is negligible. This model predicts 
an initial increase of $H_{eb}$ with $J_{eb}$ for small values of  $J_{eb}$, followed by a saturation for large enough 
values of  $J_{eb}$ at
\begin{equation}
\label{eq3}
H_{EB} = 2 \frac{\sqrt{\omega J_{A}K_{A}}}{L_{FM}},
\end{equation}
\noindent where $K_A$ is the anisotropy constant of the AFM and $\omega$ is constant depending on the lattice structure.
The previous results suggest a monotonous behavior of the bias field when the anisotropy of the FM film is negligible, 
with a linear  dependence of $H_{eb}$ with $J_{eb}$ for small values of  $J_{eb}$ and a saturation for large values of it. 
As we will show in the next section, such scenario can change substantially depending on the 
effective anisotropy of the AFM film.
\section{Microscopic model and Numerical Simulations}
\label{simulations}
\subsection{The microscopic model}

\begin{figure}
\includegraphics*[width=12cm,angle=0]{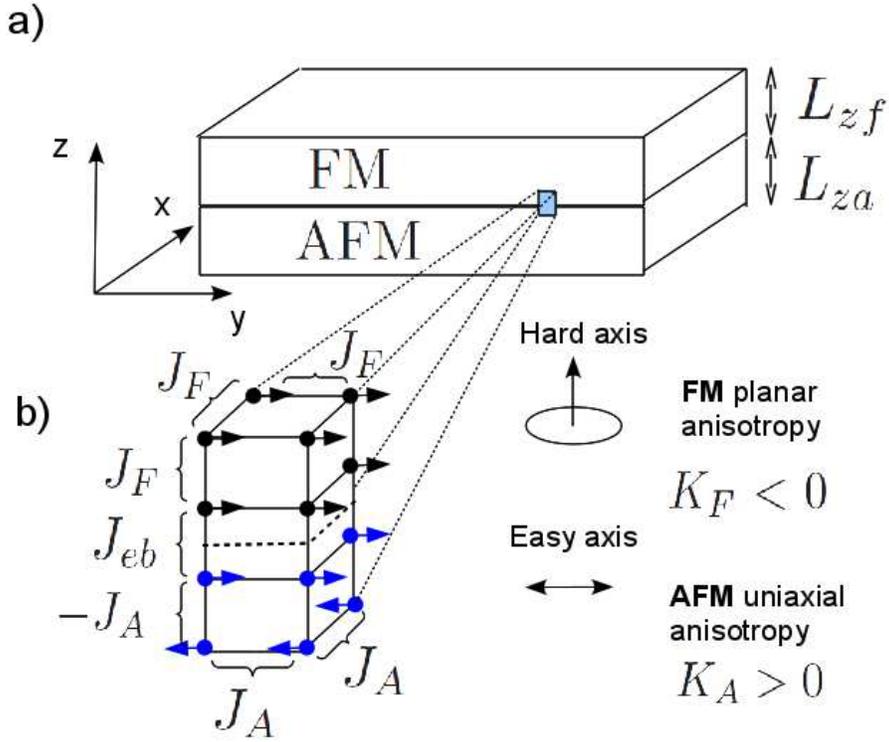}
\caption{\label{fig1} a) Scheme of the bilayer system including the reference frame used throughout this paper. 
b) Schematic picture of the system modeled by the Hamiltonian  (\ref{eq1}) in the $sc$ lattice case. Here we show
ground state configuration with the corresponding interactions.}
\end{figure}

We considered a FM film mounted over an AFM film as depicted in figure \ref{fig1}a. 
The films are magnetically coupled to each other by exchange interactions and the structure of both films 
is either $bcc$ or $sc$, assuming a perfect match across the FM/AFM interface. The system is ruled  by the 
following Hamiltonian,
\begin{eqnarray}
H &=&  -J_F\!\!\!\!\!\! \sum_{<\vec{r},\vec{r}'> \in \mbox{\footnotesize FM}} \vec{S}_{\vec{r}} \cdot \vec{S}_{\vec{r}'} 
- K_F \sum _{\vec{r} \in \mbox{\footnotesize FM}} (S^z_{\vec{r}})^2 \nonumber \\
& & - \sum _{<\vec{r},\vec{r}'> \in \mbox{\footnotesize AFM}} J_{AF}(\vec{r}-\vec{r}\,') \vec{S}_{\vec{r}} \cdot \vec{S}_{\vec{r}'}
- K_A  \sum _{\vec{r} \in \mbox{\footnotesize AFM}} (S^y_{\vec{r}})^2 \nonumber \\ 
& & -J_{eb}\!\!\!\!\!\!\!\!\!\!\! \sum _{<\vec{r},\vec{r}'> \in \mbox{\footnotesize FM/AFM}} \vec{S}_{\vec{r}} \cdot \vec{S}_{\vec{r}'}
 - h \sum _{\vec{r}} S^y_{\vec{r}}, 
\label{eq1}
\end{eqnarray}
\noindent where $\vec{S}_{\vec{r}}$ is a classical Heisenberg
spin ($|\vec{S}_{\vec{r}}|=1$) located at the node $\vec{r}$ of the lattice.
$<\vec{r},\vec{r}\,'>$ denotes a sum over nearest-neighbors
pairs of spins, $J_F>0$ is the exchange constant of the FM and $J_{AF}(\vec{r}-\vec{r}\,')$ is
the strength of the AFM exchange interactions which explicitly depends on the vector $r-r'$. 
This dependency of $J_{AF}$ on $\vec{r}-\vec{r}\,'$ is introduced in order to set an uncompensated interface 
at the AFM. For the bcc lattice we set $J_{AF}=-J_A$ with $J_A>0$ for all pairs $(\vec{r},\vec{r}\,')$, while 
for the sc lattice we set $J_{AF}=J_A$ if $(\vec{r},\vec{r}\,')$ belong to the same plane parallel to the interface 
and $J_{AF}=-J_A$ otherwise (see figure \ref{fig1}b). $J_{eb}>0$ represents the exchange coupling 
between the FM and the AFM interface layers of the films, $K_F$ is the anisotropy constant of the FM, $K_A$ is the 
AFM anisotropy and $h$ is an external homogeneous magnetic field oriented along the $y$ direction. We assumed that:
\begin{enumerate}
\item $K_F<0$ in order to ensure the FM anisotropy term tends to align
the spins on the plane of the film, mimicking the dipolar shape anisotropy, as usual \cite{Nowak02JMMM,Kim05PRB};
\item $K_A>0$ in order to introduce an uniaxial anisotropy along the $y$ direction in the AFM material \cite{Lederman04PRB},
\end{enumerate}

We carried out Monte Carlo  simulations using Metropolis algorithm and Hamiltonian (\ref{eq1}).
In our simulations $L_x$ and $L_y$ are the lateral dimensions of the films,
in units of the lattice parameter, and $L_{za}$ and $L_{zf}$ are the
thicknesses  of the FM and AFM films, respectively, measured in the
same units. The total number of spins is then $N=\eta \, L_x \, L_y \,
(L_{za}+L_{zf})$ where $\eta=1,(2)$ for the $sc$, ($bcc$) lattice.
Periodic boundary conditions were imposed in the plane of the film while
we used open boundary conditions in the perpendicular direction to the film.
For each point in the magnetization curve presented in this work, we took
$10^4$ Monte Carlo Steps per site (MCS) to thermalize the system and then the
same number of MCS to calculate the temporal averages of the magnetization.
We follow the ideas used in Refs. \cite{Wang01PRB,Billoni05PRB}, where at each spin 
actualization the direction of the spin is adjusted in a cone in such a way to maintain
an acceptation rate close to 0.46. This is an approximation to a Landau-Lifshit-Gilbert
Langevin dynamics in the high damping limit \cite{Nowak00PRL}.
We set the following dimensions for the system, $L_x=L_y=40$ and
$L_{za}=L_{zf}=12$, and fix the following parameters: $J_F=9.56J$, $J_A=-J$
and $K_F=-0.5J$, where $J$ is an arbitrary parameter that sets the
energy units.  $J_{eb}$ varies in the interval $[0,J_{F}]$
while $K_A$ can take arbitrary values. With these parameters we can emulate
a FeF$_2$-FM system in the $bcc$ lattice by choosing $K_A=1.77J$
\cite{Lederman04PRB}.
Since we are interested in the high AFM anisotropy to exchange ratio
regimen, which implies small domain wall width, the thickness of the AFM we chose is enough 
to support an AFM domain wall. On the other hand, it is known in this model \cite{Lederman04PRB,Tsai02JAP}
that for such sizes both the AFM and the FM films reach an ordered state.

\subsection{Results}
\begin{figure}
\includegraphics*[width=8cm,angle=0]{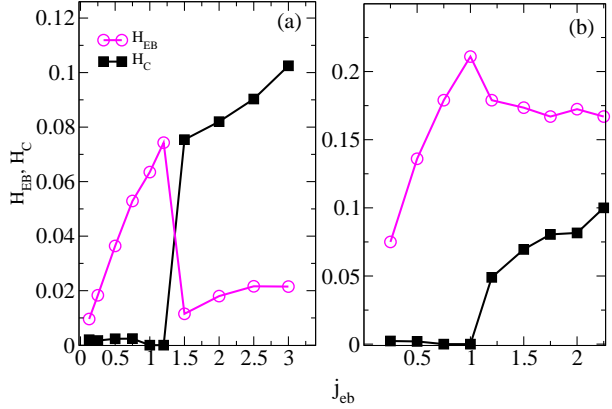}
\caption{\label{fig2} Bias field $H_{EB}$ (circles) and coercivity $H_{C}$
(squares) {\it vs.} $j_{eb}$  at $T/J_{A}=0.5$ and and $K_A/J_A=1.77$.  (a) $sc$  lattice. (b) $bcc$ lattice.}
\end{figure}

In Fig.\ref{fig2} we present the bias field $H_{EB}$ (open circles) and
coercivity $H_C$ (squares) obtained from Monte Carlo simulations as function
of the interfacial interaction strength $J_{eb}$ for the two considered
lattice structures  and for fixed values of temperature and AFM anisotropy.
 When the interfacial exchange coupling $j_{eb}\equiv J_{eb}/J_A$
is weak, $H_{EB}$ shows, for both lattice structures, a linear dependence, indicating
that the AFM spins located near the interface are fixed, and the FM film reverses
its magnetization by coherent rotation\cite{Billoni06PB}. As $j_{eb}$ increases, the
bias field reaches a maximum value at $j_{eb}^{max}$ and then abruptly drops to an almost constant value.
Notice that the drop is larger for the $sc$ lattice than for the $bcc$ one.
As it will be shown later, such effect is due to a reduction of the effective anisotropy of the AFM 
layer in the $bcc$ case.

\begin{figure}
\includegraphics*[width=8cm,angle=0]{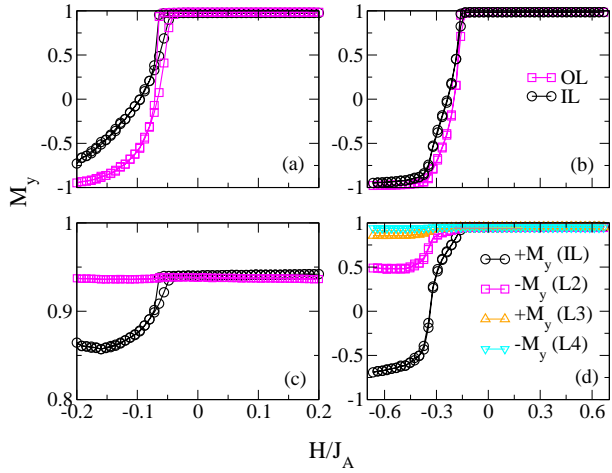}
\caption{\label{fig3}  Hysteresis loops of several atomic layers of the bilayer
 corresponding to the FM planes (top) and AFM planes (bottom) for $j_{eb}= j_{eb}^{max}$ and  $T/J_A=0.5$. 
Left panels ((a) and (c)) correspond to the  $sc$ lattice   and right panels ((b) and (d)) to the $bcc$ lattice. See text for details.}
\end{figure}

In Fig.\ref{fig3} we show the hysteresis loops of several planes of the FM
and AFM  films. These loops were obtained at $j_{eb}=j_{eb}^{max}$, just before the
drastic drops observed for $H_{eb}$ in Fig.\ref{fig2}, where the exchange bias effect
is more pronounced and the cycles are still almost reversible.

In Figs.\ref{fig3}a and \ref{fig3}b we present the magnetization in the interfacial (IL) and outer (OL) atomic
layers of the FM film. These results show that, in the two lattices, the spins rotate almost
coherently.
In Figs. \ref{fig3}c and \ref{fig3}d we show the loops of the four AFM layers nearest to the interface 
(Ln stands for the n-th atomic  layer).
Comparing the behavior in both structures we see that the $sc$ lattice is more flexible than the
$bcc$ inside the FM, but more rigid inside the AFM, because the effective anisotropy in the $sc$ is larger.
In particular, in the AFM film  of the $sc$ (Fig.\ref{fig3}c) only the first layer feels the effect of the
FM film. In the $bcc$ (Fig.\ref{fig3}d) we clearly see the formation of
a quasi-domain wall.

\begin{figure}
\includegraphics*[width=8cm,angle=0]{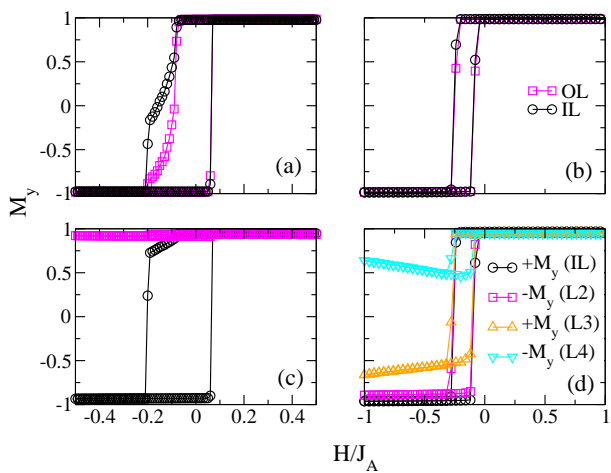}
\caption{\label{fig4}  Hysteresis loops of several atomic layers of the bilayer
 corresponding to the FM planes (top) and AFM planes (bottom) for $j_{eb}=2$ and  $T/J_A=0.5$. Left panels ((a) and (c)) correspond
to the  $sc$ lattice   and right panels ((b) and (d)) to the $bcc$ lattice.}
\end{figure}

In Fig.\ref{fig4} we plot the same quantities as in Fig.\ref{fig3}  for
a value of $j_{eb}$ above the peak, where the bias field has
already diminished abruptly. Unlike the previous case (Fig.\ref{fig3}),  here the
AFM layers show hysteresis behavior for both the $sc$ and the $bcc$
lattices (Figs.\ref{fig4}c and \ref{fig4}d respectively). This indicates
that the drop in the bias field is associated with the onset of
irreversible changes in the magnetic dynamics. As occurred below
the peak (Fig.\ref{fig3}) the changes in the AFM are constrained
to the first planes near the interface. It is worth stressing
that now the hysteresis phenomenon also appears in the AFM layers, as evidenced
in the behavior of the coercivity in Fig.\ref{fig2}.

\begin{figure}
\includegraphics*[width=8cm,angle=0]{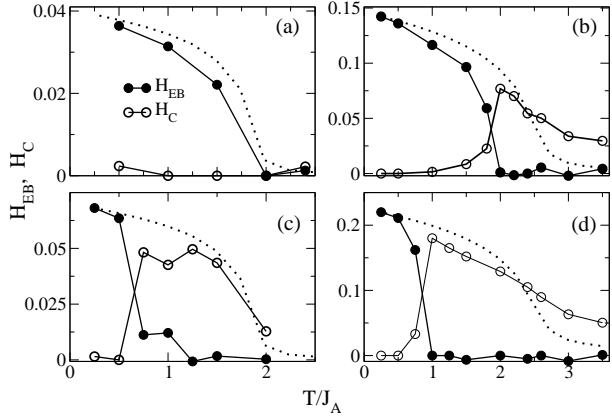}
\caption{\label{fig5}  $H_{EB}$ and $H_{C}$ {\it vs} $T/J_A$
for $K_{A}/J_A=1.77$ and two interfacial exchange interactions: top  $j_{eb}=0.5$ ((a) and
(b)) and bottom $j_{eb}=1$ ((c) and (d)). Left $sc$ ((a) and (c))
and right $bcc$  ((b) and (d)) lattices. The dotted lines represent the staggered magnetization
of the AFM at zero external magnetic field, normalized to the value of the bias field
at the lowest temperature.}
\end{figure}
\begin{figure}
\includegraphics*[scale=0.35,angle=0]{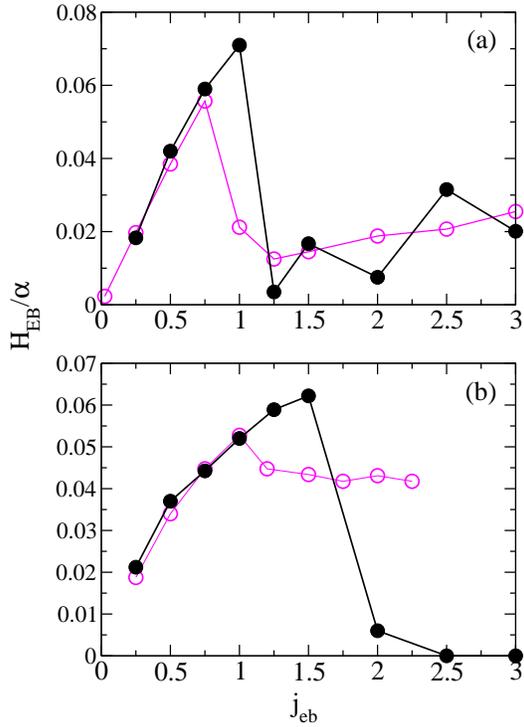}
\caption{\label{fig6} $H_{EB}/\alpha$  {\it vs} $j_{eb}$ for $T/T_c = 0.14$ and different values of $K'=K_A/\alpha J_A$.
 Open symbols: bcc lattice. Filled symbols: sc lattice. (a) $K'=1.77$. (b) $K'=0.4425$.}
\end{figure}

Next we analyzed the  temperature dependence of the overall magnetic behavior. 
In Fig.\ref{fig5} we present the bias field $H_{EB}$ and
the coercivity $H_{C}$ as a function of temperature for two values
of the interfacial interaction strength: $j_{eb} = 0.5$ (Figs.\ref{fig5}a and
\ref{fig5}b) and $1.0$ (Figs.\ref{fig5}c and \ref{fig5}d). The left
panels  correspond to the $sc$ lattice and the right
ones  to the $bcc$ lattice. The dotted lines
represent the staggered magnetization of the AFM at zero external
magnetic field, normalized with respect to the value of the bias field at the
lowest temperature. In Figs.\ref{fig5}b, \ref{fig5}c and \ref{fig5}d we observe
that the system presents a blocking temperature $T_B$ separating
two phases each with different magnetic behavior. At low temperature
the system is characterized by the presence of exchange bias and
almost zero coercivity. On the other hand, for $T_B < T < T_N$ the
bias disappears and the $H_C$ increases and further decays
following the behavior of the normalized staggered magnetization.
A complete different behavior is observed in Fig. \ref{fig5}a,
where the ordered phase coincides with the bias regime and therefore $T_B=T_N$. In this
case we do not observe any trace of coercivity in the simulations.
Note that for both structures, $sc$ and $bcc$, the blocking
temperature decreases with the interfacial interaction
strength, indicating that the energy barrier for de-pinning the
partial domain wall decreases as the wall approaches to a $180^{o}$
domain wall.

Finally we explored the effect of the lattice structure on the bias
field. The main difference between both lattice structures is the number of nearest neighbors  belonging to
adjacent layers of any site in the AFM film, which is four times larger in  the
$bcc$  than in the $sc$ structure. Hence, one would expect the effective anisotropy to be reduced by a factor of 4  in the $bcc$
lattice respect to the $sc$ one, while the bias field is expected to be four times larger in the $bcc$ than in the $sc$.
To check this hypothesis we calculated the bias field as a function of $j_{eb}$ in both lattices for the same value of
$K'=K_A/\alpha J_A$, with $\alpha=4$ for the $bcc$ lattice and $\alpha=1$ for the $sc$ lattice.
In order to compare the results, one has to take into account that the Curie temperature is different for
both lattice structures. Hence, both calculations were carried out keeping $T/T_c$ constant.
In Fig.\ref{fig6} we plot $H_{eb}/\alpha$ as a function of
$j_{eb}$ for high and low values of $K'$.   For large enough values of the anisotropy the previous conjecture is verified, 
namely, the only effect of changing the lattice structure is a rescaling of the bias field and the effective anisotropy. 
For small values of the anisotropy such scaling is observed as long as no hysteresis effects appear, namely, for small enough 
values of the coupling $j_{eb}$. For large values of $j_{eb}$ the bias field exhibits only a small drop  and it saturates at a 
constant value in the $bcc$ lattice, but it drops to zero in the $sc$ case. We observed that such large drop is due to the 
depinning of the quasi-domain wall, i.e. to a complete reversion of the staggered magnetization at the AFM film. This effect 
is not observed in the $bcc$ lattice (at least for the range of parameter values analyzed here). It is due  to a reduction in 
the in plane magnetization component at the AFM layers, associated with a canting of the spins which enhance the pinning of the wall.

\section{Layered model}
\label{model}

As we have seen in the previous section, the behavior of the bias field is strongly determined by the magnetization
dynamics of the atomic layers close to the interface. Moreover, we observed that, for reasonably large values of
the anisotropy the spins in each layer rotate almost coherently under the application of an external field parallel
to the interface. On the basis of these observations, we introduced a generalization of Meiklejohn-Bean \cite{Meiklejohn62JAP}
model that explicitly includes the contribution of the AFM layers close to the interface. We consider that only the $n$
layers of the AFM film closest to the interface are free to move, while the rest of the AFM layers keep the equilibrium
antiferromagnetic configuration of the bulk  at temperature $T$. Let $\vec{S}$ and $\vec{\sigma}_j$ be the average magnetization 
per layer per unit area at the FM  and the AFM j-th layers respectively. $\vec{S}$ and $\vec{\sigma}_j$ ($j=1,\ldots,n$) are 
assumed to be unit vectors parallel to the interface. The magnetization per unit area of the whole FM film is then given 
by $L_{FM}\vec{S}$ (with $L_{FM}$ the FM film thickness), since we are assuming a coherent rotation of  the whole FM film. 
The n-th layer is the closest one to the interface. We assume that the applied field $\vec{H}$ is parallel to the interface 
and only interacts with the FM film. This approximation is  valid as far as the applied field is small enough 
compared with the molecular field of the AFM. Finally, we consider the anisotropy of the AFM to be much larger than the FM one, 
so the latter can be  neglected. The Hamiltonian of the model is then given by
\begin{eqnarray}
{\cal H}_n &=& - K_A \, \sum_{i=1}^{n} \left( \sigma_i^y \right)^2 + (-1)^n \alpha J_A \sigma_0(T) \sigma_1^y 
+ \alpha J_A \sum_{i=1}^{n-1} \vec{\sigma}_i . \vec{\sigma}_{i+1} \nonumber \\
& &-J_{eb} \, \vec{\sigma}_{n} . \vec{S} -  \vec{H}' . \vec{S} \label{Hn},
\end{eqnarray}
\noindent where $\vec{H}'=L_{FM}\vec{H}$, $\alpha=4$ ($\alpha=1$)  for the $bcc$ ($sc$) lattice and  $\sigma_0(T)$ is the 
average equilibrium magnetization per unit area of one layer in the AFM bulk, assumed to be parallel to the easy axis. 
The $(-1)^n$ factor  in the second term of Eq.(\ref{Hn}) ensures the correct equilibrium configuration at zero temperature 
and magnetic field with the $n-1$ AFM  spin aligned with the FM spin. The model is then equivalent to  a  $n+1$-spin chain, 
where the first spin in the chain is subjected to a local effective field produced by the ordering in the AFM bulk, while the 
spin located at the end of the chain ($\vec{S}$) represents the FM film which interacts with an external magnetic field
and is  ferromagnetically coupled to the $n^{th}$ AFM spin.


At $T=0$ the sublattice magnetization in the bulk is saturated, so we have $\sigma_0(T)=1$. In a first approximation we can 
consider the simplest case of only one AFM layer free to move $n=1$ (see fig. \ref{fig7}a), which is enough to illustrate the 
general mechanism. The energy is then given by

\begin{equation}
\label{eq4}
E = - K_A \left( \sigma^y \right)^2 - \alpha J_A\; \sigma^y - \alpha J_{eb}\, \vec{\sigma} . \vec{S} -  \vec{H}'.\vec{S},
\end{equation}

\noindent where $\vec{\sigma} \equiv \vec{\sigma}_1$. The FM and AFM spins can be expressed in term of the angles  $\phi$ and $\theta$ 
respect to the easy axis direction $y$ of the AFM, in our case the field cooling direction (Fig.\ref{fig7}b). Then

\begin{eqnarray}
\label{eq5}
E &=& - K_A \cos^2 \phi  - \alpha J_{A}\cos \phi  - \alpha J_{eb}\cos(\theta - \phi) \\
  & & - H' \cos (\theta - \gamma), \nonumber
\end{eqnarray}
\noindent where the angle  $\gamma$ gives the applied field direction (Fig.\ref{fig7}b). From now on, we will consider the applied 
field parallel to the easy axis direction ($\gamma=\pi$). In order to obtain the hysteresis loops and the bias field, the model 
is analyzed using  standard procedures (see {\it e.g.} Ref.\cite{Geshev00PRB}). First, we equal to zero the partial derivatives 
$\partial_{\theta} E$ and $\partial_{\phi} E$ in order to obtain the  critical points:

\begin{figure}
\includegraphics*[width=12cm,angle=0]{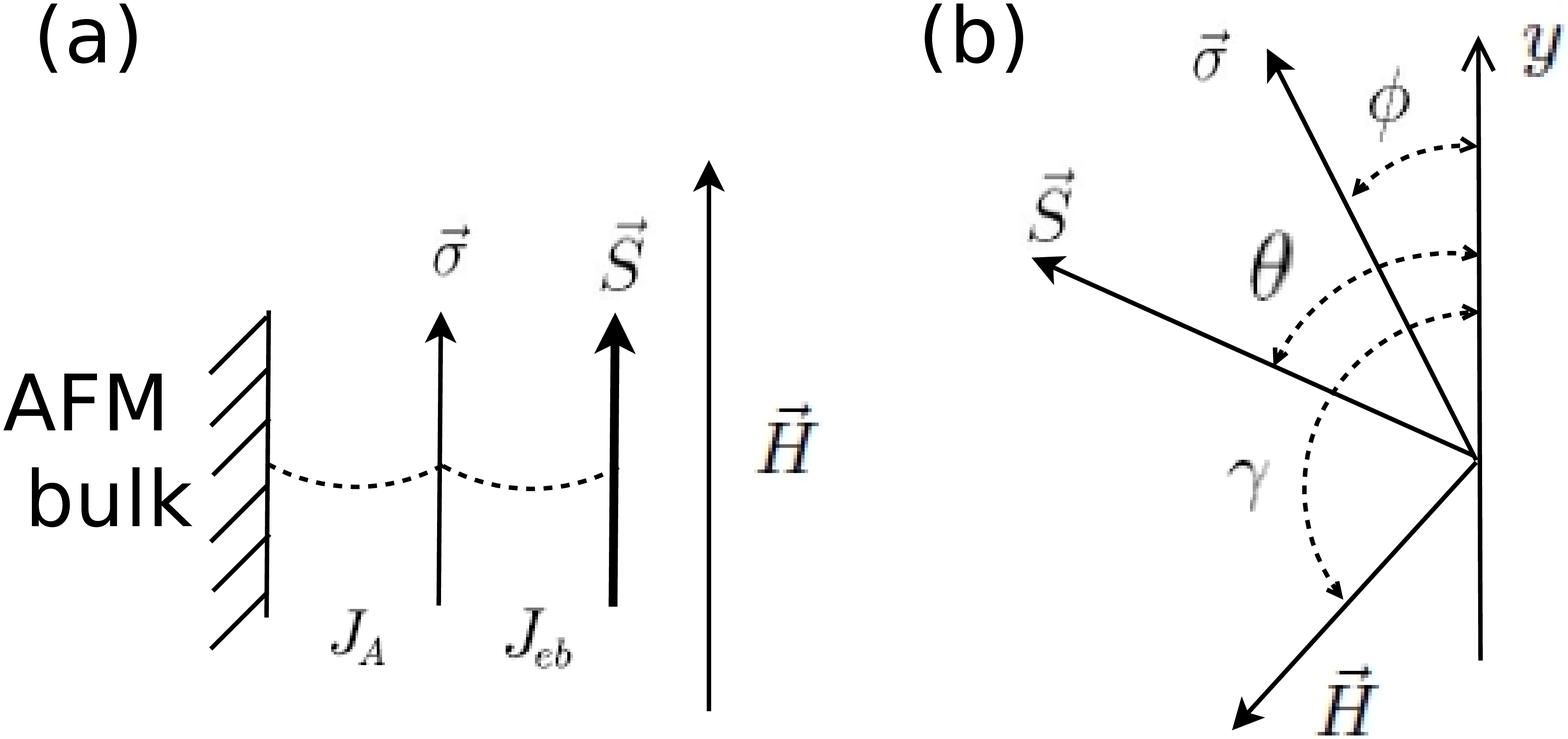}
\caption{\label{fig7} (a) Scheme of the model for $n=1$ (Eq.(\ref{eq4})). (b) Angles representing the state of the system. }
\end{figure}

\begin{eqnarray}
\label{eq6}
0 & = & \alpha J_{eb} \sin(\theta - \phi) - H' \sin(\theta) \\
        0 & = &  -\alpha J_{eb} \sin(\theta - \phi)  + \alpha J_{A} \sin{\phi} + K_A \sin(2 \phi) \nonumber
\end{eqnarray}
\noindent and then we look at the stability criteria,
$\partial_{\theta \theta}e \, \partial_{\phi \phi}e - \partial_{\theta \phi}e^2 > 0$
and $ \partial_{\theta \theta}e > 0$ to decide whether is a minimum or not. It turns out that,
\begin{eqnarray}
\label{eq7}
0 & < & \alpha J_{eb} \cos(\theta - \phi) - H' \cos \theta \\
0 & < & \alpha J_{eb} \cos (\theta - \phi) [\alpha J_{A} \cos \phi +  2 K_A \cos 2\phi] \nonumber \\
  &   &-H' \cos \theta [\alpha J_{eb} \cos (\theta - \phi) + \alpha J_A \cos \phi + 2 K_A \cos 2\phi] \nonumber
\end{eqnarray}

\begin{figure}
\includegraphics*[scale=0.29,angle=0]{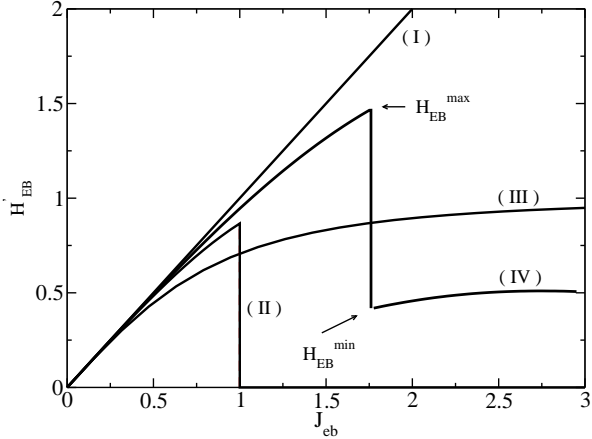}
\caption{\label{fig8}  Bias field $H'_{EB}$  as function
of interfacial exchange constant $J_{eb}$ for different values of  $J_A$ and $K_A$. (I) $J_A=0$ and $K_A \gg 1$ (Eq.(\ref{eq8}));
(II) $J_A=0$  and $K_A=1$ (Eq.(\ref{eq8})); (III) $J_{A}=1$ and $K_{A}=0$ (Eq.(\ref{eq9}));
(IV)  $J_{A}=1$ and $K_{A}=1$.}
\end{figure}

For  $J_A=0$  we recover to the  Meiklejohn-Bean  model (see Ref.[\cite{Radu08}] and references therein) and the bias 
field is given by,
\begin{equation}
\label{eq8}
H'_{eb} = \alpha J_{eb} \sqrt{1-\left(\frac{J_{eb}}{2 K_A}\right)^2},
\end{equation}
\noindent  provided that $J_{eb}<K_A$. In this range of $J_{eb}$  the coercivity field is zero. For  $J_{eb}/K_{A}>1$  the bias 
field drops to zero, while the coercivity jumps to a finite value (See Fig.\ref{fig8}), due to the complete reversal of all the spins 
in the AFM film. In the limit $K_A=\infty$ Eq.(\ref{eq8}) predicts a linear increase of $H'_{eb}$ with $J_{eb}$ (See Fig.\ref{fig8}). 
This case sets an upper limit  for the bias field of any model with uncompensated interface.

For $K_A=0$ ($J_A \neq 0$)  the coercivity is always zero and the bias field is given by

\begin{equation}
\label{eq9}
H'_{eb} =\alpha \frac{J_{A}J_{eb}}{\sqrt{J_{eb}^2+J_{A}^2}}.
\end{equation}
\noindent This equation is valid for any value of $J_{eb}$, showing a saturation at $H'_{eb}=\alpha\, J_A$ for large
values of it (See Fig.\ref{fig8}). Eq.(\ref{eq9}) becomes equivalent to the MSBK-model bias field  with zero anisotropy at the FM film, 
if we replace $J_A$ by the partial domain wall energy, namely $J_A \to 2\sqrt{K_A J_A}$. In the general case when $K_A \neq 0$ the 
coercivity is non zero and the problem has to be treated numerically.

To understand the general behavior of the bias field as a function of $J_{eb}$ let us first analyze  the structure of the energy 
landscape given by Eq.(\ref{eq5}) in the absence of  external magnetic fields. Suppose that the system was cooled under the presence 
of an external field $H_{CF}$ pointing to the positive $y$ direction. Then, the energy has an absolute minimum, corresponding to both 
magnetic variables $\vec{S}$ and $\vec{\sigma}$ pointing to the positive $y$ direction. We denote this minimum by ($\uparrow,\uparrow$). 
If the anisotropy is weak, $K_A < J_A/2$, this minimum is unique. When $K_A > J_A/2$ a second (local) minimum appears corresponding to 
both variables $\vec{S}$ and $\vec{\sigma}$ pointing to the negative $y$ direction. We denote this minimum by ($\downarrow,\downarrow$). 
If $K_A \gg J_A$ the energy difference between both minima is $\Delta E \approx 2J_A$.

Let us consider now the descending branch of an hysteresis cycle, that is, we saturate the sample with an external 
field pointing  to the positive $y$ direction  and decrease the field at regular steps until  the sample is saturated in the 
opposite direction. The effect of the inverse applied field 
on the magnetic configuration depends on the relative strength of $j_{eb}=J_{eb}/J_A$. If  $j_{eb} \ll 1$, the FM layer aligns with the 
field  when $h\equiv H'/J_A \sim j_{eb}$ but the AFM layer still points up, that is, the lower minimum ($\uparrow,\uparrow$) changes its 
configuration to  ($\uparrow,\downarrow$).  Therefore, $h_{eb} \sim j_{eb}$. When $j_{eb} \sim 1$ (and therefore  $h_{eb} \sim 1$), 
the second minimum corresponding to the  ($\downarrow,\downarrow$) configuration becomes absolute. As $j_{eb}$ further increases, 
the configuration ($\uparrow,\downarrow$) remains as a local minimum, until above certain value of $j_{eb}$ it losses stability.
Hereafter we will consider $\alpha=1$ ($sc$ lattice) for simplicity.

The typical behavior of the bias field for finite values of $K_A$ and $J_A$  is illustrated in Fig.\ref{fig8}. For low values 
of $J_{eb}$ the bias field shows a monotonous behavior, taking values between those given by Eqs.(\ref{eq8}) and (\ref{eq9}). 
At this regime, the local minimum ($\uparrow,\downarrow$) of the energy remains stable and the AFM layer forms a reversible  
quasi-domain wall, without inversion of its magnetization.  Above some maximum value $J_{eb}^{max}$, the local minimum losses 
stability  giving rise to an irreversible inversion of the AFM layer magnetization and the system   exhibits finite coercivity 
and a sudden drop in the bias field. However, at odds with the $J_A=0$ case, the bias field drops to a finite value, after 
which it increases again monotonously with $J_{eb}$ (in agreement with the simulation results of the previous section), due to 
the competition between the anisotropy and interaction of the AFM layer with the AFM bulk magnetization. For large enough values 
of $J_{eb}$ the bias field saturates into a smaller value than the $K_A=0$ case ($H_{eb}'\approx  J_A$). As $K_A$ increases 
both the drop in the bias field, as well as the value of $J_{eb}$ where it happens increase.
%

\begin{figure}
\includegraphics*[scale=0.29,angle=0]{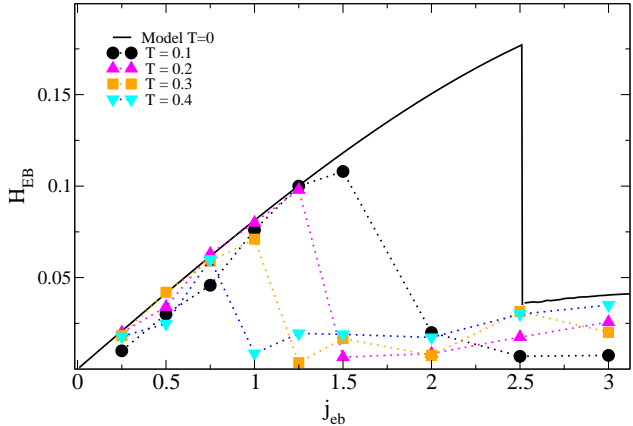}
\caption{\label{fig9}  Bias field $H_{EB}$ {\it vs.} $j_{eb}$  for different values of the temperature 
in the $sc$ lattice the anisotropy value $K_A/J_A=1.77$.}
\end{figure}

Next we compared the predictions of the model with the Monte Carlo results. In Fig.\ref{fig9} we illustrate the typical 
behavior for large values of the anisotropy. We see that, as temperature fluctuations decrease, the maximum in the bias field as 
well as the value of $j_{eb}$ where it occurs increases, due to thermal activation. Of course, this depends on the time scales 
involved in the hysteresis loop, i.e., on  the rate of variation of the field. If the rate of variation of the field is kept constant, 
the Monte Carlo results systematically approach the behavior predicted by the model as the temperature decreases, because the 
characteristic activation time systematically increases.

\begin{figure}
\includegraphics*[scale=0.29,angle=0]{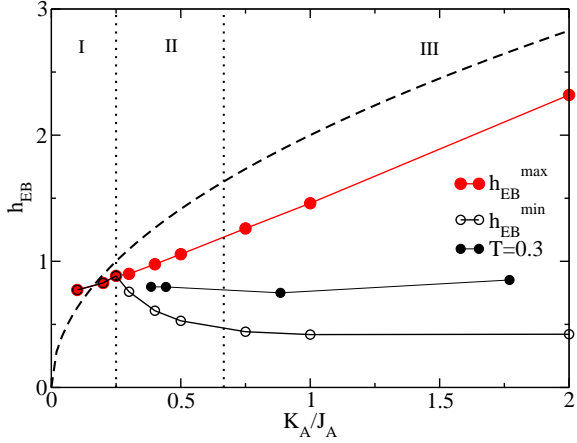}
\caption{\label{fig10}  Reduced  bias fields $h_{EB}^{max}$ and $h_{EB}^{min}$ as a function
of the reduced anisotropy $K_A/J_A$.  The dashed line is given by the MSBK-model ($h_{EB} = 2\sqrt{K_A/J_A}$)}
\end{figure}

The range of anisotropy values for which the present approximation applies can be estimated  
as $K_A/J_A >\frac{2}{3}$ since it is known in this range the domain wall width is equal to one
lattice parameter \cite{Barbara73LJDP}. When the anisotropy decreases, the domain wall thickness increases 
and more layers have to be considered  for a proper description. For small enough  values of the anisotropy 
a smooth domain wall is expected, so the behavior of the system 
should be well described by the MSBK model. 
The crossover to the regime of the MSBK model behavior can be estimated as the point 
where the energy of the domain wall equals the exchange energy of the AFM, namely $2\sqrt{K_A J_A} = J_A$, which corresponds to 
$K_A/J_A = 0.25$. This is illustrated in Fig. \ref{fig10}, where we compare the maximum bias field $h_{EB}^{max}=H_{EB}^{max}/J_A$ and 
the minimum after the drop $h_{EB}^{min}=H_{EB}^{min}/J_A$ (see Fig.\ref{fig8}) with the bias field predicted by MSBK model 
$h_{EB} = 2\sqrt{K_A/J_A}$. The vertical dotted lines divide the graph in three regions of qualitatively different behavior. The region 
of validity of the present model ($K_A/J_A>\frac{2}{3}$) is marked as  III.  In this region a quasi-domain wall forms and, unlike 
for the continuous approximation where the internal domain  wall spins change their orientation in a reversible way, now these spins can 
have an irreversible or hysteretical behavior, like when defects are present in the AFM \cite{Kim05PRB}.

In region I the continuous approach assumed in the  MSBK-model is valid. In region II neither the present model nor MSBK model are 
expected to be valid, since  the micromagnetic approach fails because the magnetization profile is no smooth in the atomic scale,  
but more than one interfacial plane is involved  in the  magnetization  process at the interface. Moreover, we have seen from the  
Monte Carlo simulations that in this region lattice structure effects can be very important. At the crossover point $K_A/J_A=0.25$ 
we see that  $h_{EB}^{max}= h_{EB}^{min}$, i.e., hysteresis disappears as expected. It is worth noting that  size effects
in the AFM become relevant only in regions I and II, in particular when the thickness of the AFM is comparable to the domain wall size.
%


Let us analyze thermal effects in the bias field when the AFM domain wall is pinned i.e., it is not
able to propagate in the bulk of the AFM material. 
Suppose that the rate of variation of the field is small enough so that the system can be assumed at thermodynamical equilibrium 
at every step of the loop. Then, the  equilibrium behavior can be obtained by computing the partition function
\begin{eqnarray}\label{Zn}
 {\cal Z}_n &=&   \int_0^{2\pi} d\phi_1 \cdots \int_0^{2\pi} d\phi_{n}\, e^{ \beta \left( K_A \, \sum_{i=1}^{n} cos^2\phi_i - 
(-1)^n \sigma_0(T) cos\, \phi_1 - \sum_{i=1}^{n-1} cos\left( \phi_i -\phi_{i+1}\right)\right)}\nonumber \\ 
& &\times \int_0^{2\pi} d\theta\, e^{\beta\, \vec{S}.\vec{\omega}},
\end{eqnarray}
\noindent where we have taken $\alpha J_A=1$, $\beta=1/k_B T$, $\phi_i$ and $\theta$ are the angles with respect
to the $y$ axis of the $i-th$ AFM and the FM spins respectively and $\vec{\omega}\equiv  \vec{H}'+ J \vec{\sigma}_{n}$ 
($J\equiv \alpha\, J_{eb}$). We assumed that the bulk AFM magnetization per layer is given by the mean field approximation \cite{Scholten05PRB,Nascimento09PRB}, namely

\begin{equation}
\label{lang}
\sigma_0(T)={\cal L}(z \beta\sigma_0(T)),
\end{equation}

\noindent where ${\cal L}(x)$ is the Langevin function and $z$ is the number of nearest neighbors which depends of the
lattice structure  $z=6,\,(8)$ for the $sc,\,(bcc)$.

The last integral in Eq. (\ref{Zn}) can be easily solved obtaining the general expression (aside from an irrelevant multiplicative factor):
\begin{eqnarray}\label{Zn2}
 {\cal Z}_n &=&   \int_0^{2\pi} d\phi_1 \cdots \int_0^{2\pi} d\phi_{n}\, e^{ \beta \left( K_A \, \sum_{i=1}^{n} cos^2\phi_i - 
(-1)^n \sigma_0(T) cos\, \phi_1 - \sum_{i=1}^{n-1} cos\left( \phi_i -\phi_{i+1}\right)\right)}\nonumber \\ 
& &\times I_0\left(\beta\, \omega(\phi_{n}) \right),
\end{eqnarray}
\noindent where $I_\nu(x)$ is the modified Bessel function and $\omega(\phi) = \sqrt{H'^2 + J^2 + 2\, J\   H' cos\, \phi}$.
%
%
\noindent The average magnetization in FM layer can be obtained as $m^F \equiv \left< cos\, \theta \right> = \frac{1}{\beta\,{\cal Z}_n } \frac{\partial {\cal Z}_n }{\partial  H'}$ and
%
%
%
%
%
%
the magnetization at the $j-th$ AFM layer, $m^{AF}_j \equiv \left< cos\, \phi_j \right>$ can be computed in a similar way.
Solving numerically the previous equations as function of the applied field and temperature we obtained the dependency of the bias field on
temperature. We considered the cases $n=1$ and $n=2$. No qualitative differences were observed. We present here the results 
for $n=1$, which are adequate for illustrating the general behavior.

\begin{figure}
\includegraphics*[width=8cm,angle=0]{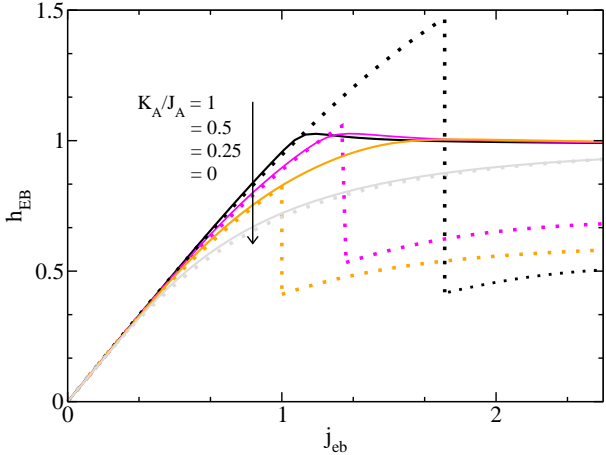}
\caption{\label{fig11} Reduced  bias field $h_{EB}$ {\it vs.} $j_{eb}$  for  different values of  $K_A/J_A$. Full lines: equilibrium curves for $n=1$ at $T/J_A=0.1$.
Dotted lines:  $T/J_A = 0$.}
\end{figure}

In Fig.\ref{fig11} we compare the equilibrium reduced bias field  $h_{EB}$ as function of $j_{eb}$  
at low temperatures (full lines) with the zero temperature curves obtained from Eq.(\ref{eq4}) (dotted lines)
for several anisotropy values. One can see that for low interfacial interaction strength $j_{eb} \ll 1$
the temperature has little effect on the bias field. In both cases an increase in the anisotropy enlarge the range of the linear behavior 
expected in the strong anisotropy limit (see fig. \ref{fig8}). This can be easily understood if we recall that in this regime the system 
behaves reversibly even at zero temperature. In other words, in both cases the behavior of the system is governed by the absolute minimum 
of the energy, so the relation $j_{eb} \sim h_{eb}$ still holds, no matter the value of the anisotropy is.

The main difference appears for  high values of $j_{eb}$. First of all,
the drop in $h_{EB}$ observed in the $T=0$  curves is absent in the thermalized curves, since of course at equilibrium there is no 
coercivity. Second, the bias field $h_{EB}$ saturates to the value $h_{EB} \sim 1$ as $j_{eb}$ increases ($j_{eb}>1$) independently of 
the anisotropy, contrasting with the  $T=0$ curves where the maximum value of $h_{EB}$ increases with the anisotropy.
When $j_{eb} \gg 1$ the applied field changes the relative depth of  the two energy minima. When $h \sim 1$  the two minima have the 
same energy and the magnetization at the FM layer inverts $m^F=0$, independently of the anisotropy. Therefore,  $h_{eb} \sim 1$, i.e. 
the bias field reaches the saturation value observed in fig. \ref{fig11}. On the other hand, the  bias field  at zero temperature 
continuously grows with the anisotropy due to the fact that the energy barriers between the minima increase with the anisotropy.

It is worth remarking that, even at equilibrium, the bias field exhibits a maximum at $j_{eb} \sim 1$ for large values of the anisotropy.

\section{Discussion}
\label{conclu}

We found that, in fully uncompensated interfaces, the bias field displays a non monotonous dependence  on the interfacial 
interaction strength. 
Depending on the temperature and on the anisotropy to exchange  ratio $K_A/J_A$ of the AFM, $H_{EB}$ can present a peak 
as a function of $J_{eb}$.  In particular, the peak is observed  at low temperatures and high enough ratios $K_A/J_A$.  
When it is present, the peak position  moves toward lower values of $J_{eb}$ as the temperature is increased, while below a certain 
temperature (low compared with the blocking temperature) the peak disappears. 
The peak is associated with the onset of coercivity, i.e., with the appearance of hysteresis for large values of $J_{eb}$.

When $K_A/J_A>\frac{2}{3}$ (region III in Fig.\ref{fig10}), the behavior of the bias field is completely determined by the 
dynamics of the interfacial AFM layer.  For low values of $J_{eb}$ the interfacial layer rotates coherently  forming a quasi-domain 
wall that changes reversibly  with the applied field. 
In this regime the bias field increases almost linearly with $J_{eb}$ and thermal effects are negligible. Above a certain critical 
value of  $J_{eb}$ the quasi domain wall loses stability and the magnetization of the interfacial AFM layer changes irreversibly. 
In other words, the bias field is  controlled by the stability of the interfacial layer. This scenario,  supported by both 
the Monte Carlo simulations and the simple layered model 
introduced here, explains why the bias field can be drastically reduced by increasing the interfacial interaction strength (Fig. \ref{fig2}). 
Also in this regime ($K_A/J_A>\frac{2}{3}$), the behavior of the bias field is  independent of the lattice structure. 
In other words, a change in the crystalline  structure is just equivalent to a rescaling of the effective anisotropy of the AFM.

When  $K_A/J_A< \frac{2}{3}$, the system can  still exhibit hysteresis and a peak in the bias field (region II in Fig.\ref{fig10}), 
but the the width of the domain wall increases as $K_A/J_A$ decreases. In this case bias, field is controlled by the intrinsic 
pinning due to the anisotropy. 
Namely, for large values of $J_{eb}$ the bias field reduces because of the depinning of this domain wall, which depends strongly on
the lattice structure. In particular, preliminary results showed that the pinning is stronger in the $bcc$ than the $sc$ lattice, 
due to canting  effects in the AFM layers. A detailed study of such effect is underway and will be published elsewhere.

In both regimes (II and III) the maximum bias field is smaller than the value predicted by MSBK-model.
These results offer certain insights about experimental findings in FeF$_2$ systems\cite{Moran98APL,Nogues99PRB} 
($K_A/J_A>\frac{2}{3}$), where in a fully uncompensated interface the bias field is much lower than expected. 
In particular, it becomes noticeable at very low temperatures. According  to our results, if the interfacial strength interaction 
is strong the bias field becomes no null only at very low temperatures compared with the Neel temperature of the antiferroagnet. 

Summarizing,  depending on the anisotropy to exchange ratio $K_A/J_A$ the bias field is controlled either by the 
intrinsic pinning of an extended domain wall parallel to the interface (low anisotropy regime) or by the stability of 
the first AFM  interfacial plane near the interface (sharp domain wall limit).

\section*{Acknowledgments}
This work was partially supported by grants from CONICET (Argentina), Agencia C\'ordoba Ciencia (Argentina), SeCyT,
Universidad Nacional de C\'ordoba (Argentina).

\section*{References}

\end{document}